\documentclass[letterpaper,twocolumn,10pt]{article}
\usepackage{adjustbox}
\usepackage[english]{babel}
\usepackage{makecell}
\usepackage{usenix}

\usepackage{blindtext}

\usepackage{tikz}
\usepackage{amsmath}

\usepackage{filecontents}

\usepackage{tikz}
\usepackage{xcolor}
\usepackage{xspace}
\usepackage{tikz}
\usepackage{pifont}
\usepackage[english]{babel}
\usepackage{blindtext}
\usepackage{lipsum}

\usepackage[ruled,vlined,linesnumbered]{algorithm2e}
\usepackage{booktabs}

\usepackage{filecontents}
\usepackage{xspace}
\usepackage{xcolor}
\usepackage{hhline}
\usepackage{multirow}
\usepackage{soul}

\usepackage{array}
\usepackage{comment}
\usepackage{listings}
\usepackage{amssymb}
\usepackage{dsfont}
\usepackage{algorithmic}
\usepackage{xurl}
\usepackage{setspace}
\usepackage[flushleft]{threeparttable}
\usepackage{enumitem}
\usepackage{enumerate}
\usepackage{float}
\usepackage{subcaption}
\usepackage{graphicx}
\usepackage{breqn}

\usepackage{caption}
\usepackage{ltablex}

\usepackage{ifluatex}

\usepackage{outlines}

\setlist[enumerate]{noitemsep, topsep=0pt}
\setlist[itemize]{noitemsep, topsep=0pt}

\usepackage[compact]{titlesec}
\renewcommand{\paragraph}[1]{\noindent\textbf{#1}}
\titlespacing{\paragraph}{0pt}{*0.5}{*1}
\titlespacing\section{0pt}{*1.8}{*1.1}
\titlespacing\subsection{0pt}{*1.5}{*1.1}
\titlespacing\subsection{0pt}{*1.3}{*0.8}
\usepackage{booktabs}

\begin{document}

\newcommand*{\affmark}[1][*]{\textsuperscript{#1}}
\newcommand*{\affaddr}[1]{#1}


\title{CAIP: Detecting Router Misconfigurations with Context-Aware Iterative Prompting of LLMs}
\newcommand{\eg}{{\it e.g.}}
\newcommand{\ie}{{\it i.e.}}
\newcommand{\etal}{{\it et al.}}
\newcommand{\sysname}{{CAIP}}
\newcommand{\edit}[1]{{\color{black} #1}}
\newcommand{\jc}[1]{{\footnotesize\color{orange}{(JC: #1)}}}
\newcommand{\jcedit}[1]{{\color{orange} #1}} 
\newcommand{\chase}[1]{{\color{brown}{(Chase: #1)}}}
\newcommand{\todo}[1]{{\color{red}{(TODO: #1)}}}
\definecolor{darkkhaki}{rgb}{0.74, 0.72, 0.42}
\newcommand{\sr}[1]{{\color{cyan!70!blue}{(Siddhant: #1)}}}
\newcommand{\jack}[1]{{\color{purple}{(Jack: #1)}}}
\newcommand{\aaron}[1]{{\color{teal}{(Aaron: #1)}}}

\newcommand{\smallindent}{\hphantom{N}}

\newcommand{\vspacesize}{0.2cm}

\newcommand{\fillme}{{\bf XXX}\xspace}

\newcommand*\circled[1]{\tikz[baseline=(char.base)]{
            \node[shape=circle,fill,inner sep=2pt] (char) {\textcolor{white}{\footnotesize{#1}}};}}

\newcommand{\name}{NAME\xspace}

\newcounter{packednmbr}
\newenvironment{packedenumerate}{\begin{list}{\thepackednmbr.}{\usecounter{packednmbr}\setlength{\itemsep}{0.5pt}\addtolength{\labelwidth}{-4pt}\setlength{\leftmargin}{2ex}\setlength{\listparindent}{\parindent}\setlength{\parsep}{1pt}\setlength{\topsep}{0pt}}}{\end{list}}
\newenvironment{packeditemize}{\begin{list}{$\bullet$}{\setlength{\itemsep}{0.5pt}\addtolength{\labelwidth}{-4pt}\setlength{\leftmargin}{2ex}\setlength{\listparindent}{\parindent}\setlength{\parsep}{1pt}\setlength{\topsep}{2pt}}}{\end{list}}
\newenvironment{packedpackeditemize}{\begin{list}{$\bullet$}{\setlength{\itemsep}{0.5pt}\addtolength{\labelwidth}{-4pt}\setlength{\leftmargin}{\labelwidth}\setlength{\listparindent}{\parindent}\setlength{\parsep}{1pt}\setlength{\topsep}{0pt}}}{\end{list}}
\newenvironment{packedtrivlist}{\begin{list}{\setlength{\itemsep}{0.2pt}\addtolength{\labelwidth}{-4pt}\setlength{\leftmargin}{\labelwidth}\setlength{\listparindent}{\parindent}\setlength{\parsep}{1pt}\setlength{\topsep}{0pt}}}{\end{list}}
\let\enumerate\packedenumerate
\let\endenumerate\endpackedenumerate
\let\itemize\packeditemize
\let\enditemize\endpackeditemize

\newcommand{\tightcaption}[1]{\vspace{-0.15cm}\caption{{\normalfont{\textit{{#1}}}}}\vspace{-0.3cm}}
\newcommand{\tightsection}[1]{\vspace{-0.3cm}\section{#1}\vspace{-0.2cm}}
\newcommand{\tightsectionstar}[1]{\vspace{-0.17cm}\section*{#1}\vspace{-0.08cm}}
\newcommand{\tightsubsection}[1]{\vspace{-0.25cm}\subsection{#1}\vspace{-0.1cm}}
\newcommand{\tightsubsubsection}[1]{\vspace{-0.01in}\subsubsection{#1}\vspace{-0.01cm}}

\newcommand{\bigO}{\mathrm{O}}
\newcommand{\twlog}{w.l.o.g.\xspac}

\newcommand{\myparashort}[1]{\vspace{0.05cm}\noindent{\bf {#1}}~}
\newcommand{\mypara}[1]{\vspace{0.05cm}\noindent{\bf {#1}:}}
\newcommand{\mysubpara}[1]{\vspace{0.05cm}\noindent{\textit{#1}}}
\newcommand{\myparatight}[1]{\vspace{0.02cm}\noindent{\bf {#1}:}~}
\newcommand{\myparaq}[1]{\smallskip\noindent{\bf {#1}?}~}
\newcommand{\myparaittight}[1]{\smallskip\noindent{\emph {#1}:}~}
\newcommand{\question}[1]{\smallskip\noindent{\emph{Q:~#1}}\smallskip}
\newcommand{\myparaqtight}[1]{\smallskip\noindent{\bf {#1}}~}

\newcommand{\cmark}{\ding{51}}%
\newcommand{\xmark}{\ding{55}}%



\definecolor{backcolour}{rgb}{0.96,0.96,0.96}
\definecolor{codegray}{rgb}{0.5,0.5,0.5}
\definecolor{deepblue}{rgb}{0,0,0.6}
\definecolor{deepred}{rgb}{0.6,0,0}
\definecolor{deepgreen}{rgb}{0,0.5,0}
\lstdefinestyle{mystyle}{
    backgroundcolor=\color{backcolour},   
    commentstyle=\color{codegreen},
    morekeywords={self, True},
    keywordstyle=\color{deepblue},
    numberstyle=\tiny\color{codegray},
    emph={MyClass,__init__,EncodingType,Image},
    emphstyle=\color{deepred},
    stringstyle=\color{deepgreen},
    basicstyle=\ttfamily\footnotesize,
    breakatwhitespace=false,         
    breaklines=true,                 
    captionpos=b,                    
    keepspaces=true,                 
    numbers=left,                    
    numbersep=5pt,                  
    showspaces=false,                
    showstringspaces=false,
    showtabs=false,                  
    tabsize=1
}

\author{
	{\rm Xi Jiang}\\
	University of Chicago
	\and
	{\rm Aaron Gember-Jacobson}\\
	Colgate University 
	 \and
	 {\rm Nick Feamster}\\
	University of Chicago
} 

\maketitle


\begin{abstract}

Model checkers and consistency checkers detect critical errors in router
    configurations, but these tools require significant manual effort to
    develop and maintain. 
    LLM-based Q\&A models have emerged as a promising 
    alternative, allowing users to query partitions of configurations 
    through prompts and receive answers based on learned 
    patterns, thanks to transformer models pre-trained on 
    vast datasets that provide generic configuration context for interpreting 
    router configurations.
    Yet, current methods of partition-based prompting often
    do not provide enough network-specific context 
    from the actual
    configurations to enable accurate inference.
    We introduce a Context-Aware Iterative Prompting (\sysname{}) framework that
    automates network-specific context extraction and optimizes LLM prompts
    for more precise router misconfiguration detection. \sysname{} addresses
    three challenges: (1) efficiently mining relevant context from complex configuration files, (2) accurately distinguishing between
    pre-defined and user-defined parameter values to prevent irrelevant
    context from being introduced, and (3) managing prompt context overload with
    iterative, guided interactions with the model.
 Our evaluations on synthetic and real-world configurations show that 
 \sysname{} improves misconfiguration detection accuracy by more than 
 30\% compared to partition-based LLM approaches, model checkers, and 
 consistency checkers, uncovering over 20 previously undetected 
 misconfigurations in real-world configurations.

\end{abstract}

\pagestyle{plain}

\begin{sloppypar}

\section{Introduction}
\label{sec:intro}

Detecting router misconfigurations is crucial for maintaining the stability,
security, and performance of network infrastructures. Whether due to
overlooked errors in existing setups or mistakes introduced during
configuration changes, misconfigurations can lead to black holes, unintended
network access, inefficient routes, and other performance and security
issues.  For example, a misconfigured access control list (ACL) may allow
unauthorized traffic, exposing the network to potential attacks; a buggy
routing policy may create loops, rendering services inaccessible and degrading
performance.

Researchers and practitioners have developed a plethora of tools for detecting
network misconfigurations. Model checkers~\cite{fogel2015general,
beckett2017general, abhashkumar2020tiramisu, prabhu2020plankton, zhang2022sre,
steffen2020netdice, ye2020hoyan, ritchey2000using,al2011configchecker,
jeffrey2009model} model a network's routing and forwarding behaviors based on
protocol semantics and device configurations, and check whether
engineer-specified reachability and resilience policies are satisfied.
Consistency checkers~\cite{kakarla2024diffy, kakarla2020finding,
le2006minerals, feamster2005detecting,
tang2021campion,le2008detecting,le2006characterization} compare configurations
within and across devices and flag inconsistencies and deviations from best
practices. LLM-based Q\&A
tools~\cite{bogdanov2024leveraging,chen2024automatic,wang2024identifying,liu2024large,
wang2024netconfeval, lian2023configuration} parse configuration files and
query pre-trained sequential transformer models through prompts to detect syntax and subtle
semantic issues.

A fundamental feature of all configuration checkers is their reliance on {\em
context}. Model checkers require protocol specifications, current (or
proposed) device configurations, and forwarding policies; consistency checkers
require configurations from multiple devices and a set of best practices; and
LLMs require a sufficient number of relevant configuration lines in query
prompts. Moreover, the correctness and completeness of this context affects 
the accuracy of any configuration checker. For example, providing an incomplete set of
forwarding policies, best practices, or configuration lines
may cause misconfigurations to be missed. Similarly, improperly modeling
protocol semantics~\cite{birkner2021metha, ye2020hoyan}, excluding certain
configurations~\cite{xu2023netcov}, or supplying unrelated configuration
lines~\cite{liskavets2024prompt,tian2024examining,khurana2024and,
shvartzshnaider2024llm} may cause false alarms.

Although some context is easy to provide---\eg, current
config\-urations---many forms of context require significant manual effort.
For example, a detailed understanding of routing protocols, their
interactions~\cite{le2007rr}, and vendor-specific nuances~\cite{ye2020hoyan}
is required to accurately model networks' routing and forwarding behaviors.
Similarly, context concerning packet and route filter semantics is required to
accurately detect inconsistencies in filter
configurations~\cite{kakarla2020finding}.

Because LLMs are typically trained on vast volumes of data---which likely
includes protocol standards, vendor documentation, and sample
configurations---a significant amount of \textit{generic configuration
context} is already embedded in large-scale pre-trained models
(PTMs)~\cite{qiu2020pre}. However, such pre-learned generic context often
needs to be supplemented with \textit{network-specific context} included in the query prompts, which details
the actual statements within the current configuration(s) under
analysis. 
Including this network-specific context in the prompt to the LLM is essential
for detecting misconfigurations that are specific to the device or network
being checked.
However, state-of-the-art 
approaches, such as feeding a device's entire configuration into a single
prompt~\cite{lican,li2024long} can result in context and token overload,
diluting the model's focus. On the other hand, partitioning a
device's configuration into multiple
prompts~\cite{lian2023configuration,chen2024automatic,wang2024identifying}
may cause the LLM to ignore dependencies between configuration lines,
resulting in failures to detect certain misconfigurations.


In this paper, we address the problem of precisely extracting and integrating network-specific context into LLM prompts to enhance the accuracy of router misconfiguration detection. We introduce the
Context-Aware Iterative Prompting (\sysname{}) framework, which
solves three key challenges in this process:

\mypara{Challenge 1: Automatic and Accurate Context Mining} Network
configurations are often lengthy and
complex~\cite{benson2009complexitymetrics}, with interrelated lines that
require careful context extraction. The challenge lies in automatically
identifying and extracting relevant context to reduce computational costs and
avoid introducing irrelevant information in prompts that could impair the accuracy of
misconfiguration detection. \textbf{Solution}: We address this challenge by
leveraging the hierarchical structure of configuration files, modeling them as
trees, where each line is a unique path from the root to a parameter value.
This representation allows us to systematically mine and adhere to three core types of
context---neighboring, similar, and referenced configuration
statements---ensuring the LLM receives the most relevant information for
accurate analysis.

\mypara{Challenge 2: Parameter Values Ambiguity in Referenceable Context}
Configurations contain both pre-defined and user-defined parameter values,
with the latter requiring context for proper interpretation. For example, an
interface might reference a user-defined access control list (ACL), which must
be interpreted based on related configuration lines. When extracting
referenced context, misidentifying these can lead to irrelevant context mining
and increased computational overhead. \textbf{Solution}: To accurately
differentiate these values, we apply an existence-based method within the
configuration tree, identifying user-defined values by their presence as
intermediate nodes in other paths. Additionally, we use majority-voting to
further resolve ambiguities, ensuring consistency in how values are treated across
different contexts.

\mypara{Challenge 3 - Managing Context Overload in Prompts}
Even with precise context extraction, the volume of relevant information can
be large to fit in a single prompt. 
Feeding excess context into LLMs can dilute relevance, reduce coherence, and degrade the model’s performance.
\textbf{Solution}: Our framework mitigates the potential for context overload
by allowing the LLM to iteratively request specific types of context, refining
the prompt based on the model’s feedback. This approach enhances the model’s
ability to focus on the most relevant information, leading to more precise
misconfiguration detection and reducing the likelihood for context overload.

We evaluate our framework through two distinct case studies.
First, using configuration snapshots from real networks, we introduce {\em synthetic} misconfigurations that mirror potential operator errors. 
We demonstrate that \sysname{} accurately infers and detects these misconfigurations, with at least 30\% more
accuracy than conventional partition-based LLM approach, model checkers, and consistency checkers.
Second, we exhaustively run \sysname{} on configuration snapshots from a medium-sized campus network, and \sysname{} is able to detect over 20 cases of previous overlooked misconfigurations.

\section{Background and Motivation}
\label{sec_background}
Network misconfigurations are a common and persistent issue, often leading to security vulnerabilities, performance degradation, or network outages~\cite{zheng2012atpg, feamster2005detecting}. Manually identifying and rectifying  misconfigurations is challenging due to the complexity and interdependency of configurations~\cite{le2007rr, benson2009complexitymetrics}.

Existing methods for automated network misconfiguration detection fall into two categories: model checking and consistency checking. Model checkers~\cite{fogel2015general, beckett2017general, abhashkumar2020tiramisu, prabhu2020plankton, zhang2022sre, steffen2020netdice, ye2020hoyan, ritchey2000using,al2011configchecker, jeffrey2009model} identify misconfigurations by deriving a logical model of the control and data planes and executing or analyzing the model to determine whether engineer-specified forwarding policies are satisfied.
For example, Batfish simulates the control and data planes under specific network conditions (e.g., link failures) and checks whether forwarding paths deviate from specified policies~\cite{fogel2015general}. Similarly, Minesweeper uses a satisfiability solver to verify whether reachability and load-balancing policies are satisfied under a range of network conditions (e.g., all single link failures)~\cite{beckett2017general}.
Although model checkers are popular, they rely on predefined rules/policies and require domain expertise to set up. Moreover, they often struggle with the nuances of particular protocols and vendors~\cite{ye2020hoyan, birkner2021metha}.
In contrast to model checkers, consistency checkers such as \textit{Diffy}~\cite{kakarla2024diffy} and \textit{Minerals}~\cite{le2006minerals} take a statistical approach by learning common configuration patterns and detecting deviations as potential errors.
For instance, Diffy learns configuration templates and identifies anomalies by
comparing new configurations to learned patterns, while Minerals applies
association rule mining to detect misconfigurations by learning local policies
from router configuration files. Unfortunately, these approaches tend to oversimplify configurations by treating deviations from standard patterns as misconfigurations, leading to false positives when valid configurations deviate for context-specific reasons.

In light of the shortcomings of these existing tools,
researchers have looked to Large Language Models (LLMs) for misconfiguration detection~\cite{bogdanov2024leveraging,chen2024automatic,wang2024identifying,liu2024large, wang2024netconfeval} due to their advanced ability to understand and process complex contextual information embedded within network configurations. In a Q\&A format, prompts containing
instructions and configuration lines are provided to the models,
which then respond to identify potential misconfigurations.
A representative tool in this category is Ciri~\cite{lian2023configuration}, a LLM-based configuration verifier.

LLMs, particularly transformer-based
models~\cite{vaswani2017attention,hill2024transformers,lin2022survey}, excel
at capturing intricate relationships between configuration elements by
leveraging self-attention mechanisms that dynamically weigh the importance of
each token (or configuration parameter) in relation to others, regardless of
their distance within the file. This allows LLMs to incorporate both local and
global dependencies, enabling them to recognize not only syntax and pattern
anomalies but also to infer potential misconfigurations based on the
underlying semantics of the configuration. The use of position encodings
ensures that the order of elements in a configuration file is considered,
allowing LLMs to assess the correctness of parameters based on their sequence.
Additionally, most LLMs rely on large-scale pre-trained models (PTMs) trained
on large amounts of data~\cite{qiu2020pre}, particularly generative pretrained
tranformers
(GPTs)~\cite{achiam2023gpt,touvron2023llama,shanahan2024talking,taylor2023galactica,brown2020language,chowdhery2023palm}.
These mechanisms enable them to obtain vast pre-learned contexts which allow
for effective generalization across various tasks.  These distinct advantages
have spurred the use of LLMs across diverse
domains~\cite{carion2020end,sheng2019nrtr,neil2020transformers,parmar2018image,chen2021developing,gulati2020conformer},
making their application in network misconfiguration detection a natural
progression.

Through this architecture and the pre-learned general network configuration contexts, LLM-based Q\&A models detect subtle errors that might arise from interactions between different configuration elements—issues often missed by conventional tools. As networks become more complex, with increasingly interdependent components, LLMs' ability to holistically analyze configurations and understand the intent behind them makes them a powerful tool for enhancing the accuracy and reliability of misconfiguration detection.
However, to fully leverage this capability, query
prompts must include all relevant context from the configuration file
when analyzing a particular configuration line or excerpt,
including related lines that might aid in detecting misconfigurations. 
Doing so helps to refine the model's response by incorporating network-specific context unique to the current configuration, rather than relying solely on pre-learned knowledge. Without this targeted context, the LLM's ability to identify potential issues is significantly diminished ~\cite{liskavets2024prompt,tian2024examining,khurana2024and, shvartzshnaider2024llm}.

\mypara{Limitations of Full-File Prompting} A na\"{i}ve approach to incorporating context would be to feed the entire
configuration file to the LLM, either all at once in a single prompt or
progressively, and let the model handle the detection. This method, while
straightforward, is highly inefficient due to the inherent token length
limitations of LLMs~\cite{xue2024repeat,yu2024breaking,gu2023mamba}. More
critically, flooding the model with all the configuration data can lead to
context overload~\cite{lican,li2024long,qian2024long}, a known issue where the
presence of excessive and irrelevant information dilutes the LLM’s capacity to
focus on important aspects of the configuration. Context overload not only
impairs the model’s performance but
can also result in a failure to detect important misconfigurations, or generating false positives
due to irrelevant context. 

\mypara{Challenges of Partition-Based Prompting} A common solution to context overload has been partition-based prompting,
where the large and complex configuration file is broken down into smaller
chunks or sections, typically based on the order in which the configuration
appears in the file. These chunks are then individually fed to the LLM for
misconfiguration detection as independent prompts. This approach reduces the
token load per prompt and ensures that the model is not overwhelmed by a
massive context. However, this method introduces a new set of problems: by
treating sections in isolation, it often fails to account for interdependent
lines that reside in different parts of the file. Network configurations,
unlike typical documents, often contain parameters that interact with or
depend on other sections that may not be adjacent in the file. While methods
like prompt chaining~\cite{wang2024identifying,bogdanov2024leveraging} guide
LLMs by linking subsequent instructions across prompts, these do not target
context chaining, leaving critical interdependencies unaccounted for.

For example, consider a configuration file where one chunk defines firewall
rules and another chunk, further down the file, defines routing policies. A
misconfiguration in the firewall rules might depend on how the routing
policies are established, but because partition-based prompting processes
these sections independently, this critical context is lost. Lost context
is particularly detrimental for detecting dependency-related
misconfigurations, where proper detection requires understanding how different
sections of the configuration work together. If the partitioned chunk only
contains locally adjacent lines that define completely different things, then
critical context from other parts of the configuration will be missing, and
the model will fail to make accurate inferences.


In this paper, we present a tool that addresses these challenges through a
context-aware, iterative prompting mechanism. This approach efficiently
manages and extracts relevant context for each configuration line under
review, ensuring the LLMs receive the necessary information in prompts without
suffering context overload. The next section outlines the challenges we faced
in developing this solution.

\section{Method}
\label{sec:method}

\begin{figure*}[t]
    \centering
    \includegraphics[width=\linewidth]{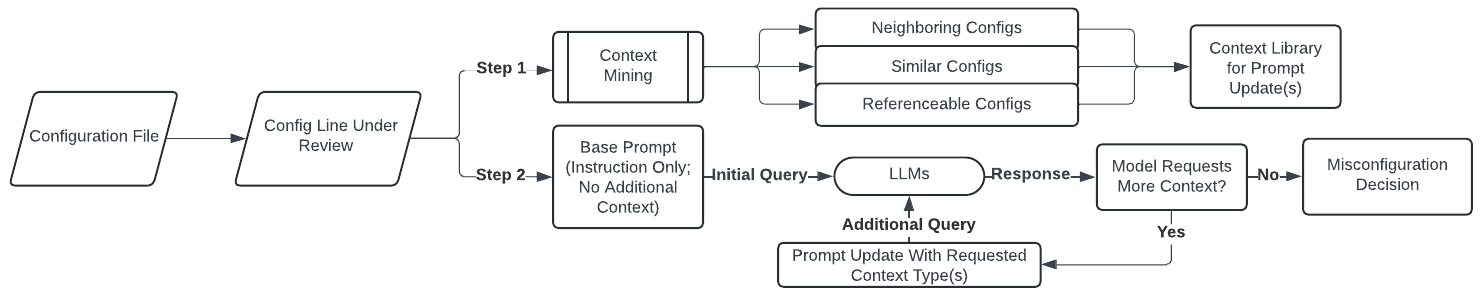}
    \caption{\sysname{} system overview.}
    \label{fig:overview}
\end{figure*}

To address the limitations of partition-based prompting,
which often fails to capture dependencies across non-adjacent sections of configuration files,
we design
Context-Aware Iterative Prompting (\sysname{}), which has two components, as
shown in Figure~\ref{fig:overview}: a context mining component and an
iterative prompting component. These components work together to automate the
process of context extraction and prompt interaction with LLMs, enabling more
accurate router misconfiguration detection.

\mypara{1. Context Mining Component} In this phase, \sysname{} mines all
relevant context for a given configuration line under examination. Relevance
here refers to configurations that, while not necessarily directly related,
provide important insights, such as neighboring configurations, similar lines
applied in different contexts, or referenceable configurations that define key
parameters.


\mypara{2. Iterative Prompting Component} Once the relevant context has been
mined, the online component engages the LLM through an iterative prompting
process. Instead of overwhelming the model with all the extracted context at
once, \sysname{} interacts with the LLM in a guided, sequential manner.

We now explain the challenges encountered in designing these two components and present the solutions.

\subsection{Context Mining}\label{mining_method} \subsubsection{Challenge 1 -
Efficient and accurate context mining} \label{challenge_1} Configuration files
are often lengthy and complex, consisting of multiple related and
interdependent lines that define various aspects of the behavior of the
corresponding network device. When
analyzing a specific configuration line, efficiently identifying
extracting the relevant context can reduce the computational burden on LLMs
and minimize the risk of introducing irrelevant information that could degrade
the accuracy of misconfiguration detection. This task is challenging because,
although these configurations are typically written in a machine and
human-readable format, automating the process of context extraction requires a
deep understanding of the hierarchical and interconnected nature of the
configuration data. For example, in a network configuration, a line that
specifies an access control rule might depend on prior lines that define
network segments or user roles. Without properly extracting and including this
context in prompt, the LLM might misinterpret the rule, leading to false
positives or missed misconfigurations.

\begin{figure}[t]
    \centering
    \fbox{\includegraphics[width=0.9\columnwidth]{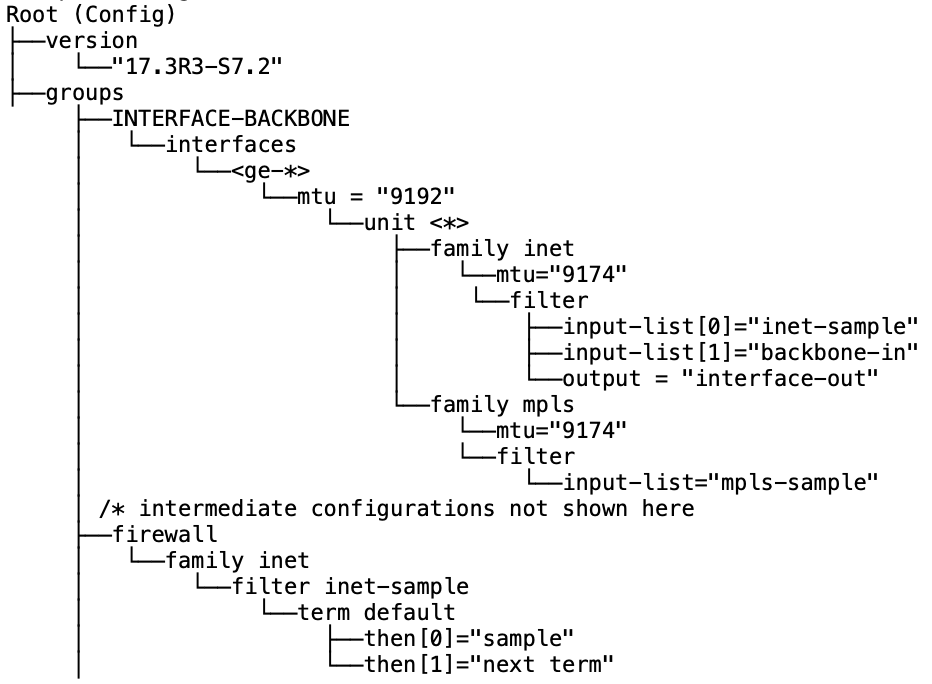}}
    \caption{Example snippet of tree-formatted Junpiter router configuration file.}
    \label{fig:tree}
\end{figure}

To address this challenge, we rely on two observations:

\mypara{1. Hierarchical Structure of Configuration Files}
    Configuration files are typically written in a structured, hierarchical format that can be effectively modeled as a tree, as shown in the example in Figure~\ref{fig:tree}. In this tree representation (\(T\)), each node (\(V\)) corresponds to a specific configuration element, and edges represent the relationships or dependencies between these elements. Let any unique path \(
P = \{ V_1 \rightarrow V_2(P) \rightarrow V_3(P) \rightarrow \dots \rightarrow V_k(P) = v_k(P) \}
\) in this tree have \(k\) nodes, where the value of \(k\) can vary depend on the path taken. The root node (\(V_1\)) represents the entry point of the configuration file (thus, it functions solely as the root of the tree, providing no semantic meaning, and remains identical regardless of the path taken), while the immediate following node (\(V_2(P)\)) corresponds to the broadest configuration category in this specific path. Intermediate nodes at subsequent levels (\( V_2(P), V_3(P), \dots, V_{k-1}(P) \)) represent increasingly nested configuration sections or subcategories, each refining the configuration context. The leaf node \(V_k(P)\) (or parameter node) represents the final configurable parameter in this path, with the associated value \(v_k(P)\), \ie, parameter values, indicating the specific configuration setting. An unique configuration line is thus a complete path in the tree, starting from the root node and terminating at a parameter node, where the path reflects the hierarchical structure and relationships defined in the configuration file.

\begin{figure}[t]
    \centering
    \fbox{\includegraphics[width=0.9\columnwidth]{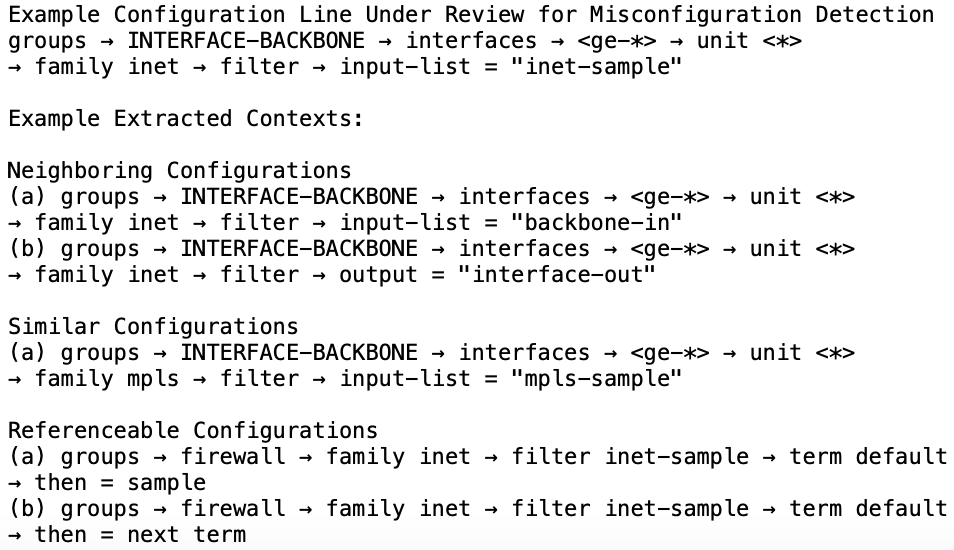}}
    \caption{Example context mined on selected config line.}
    \label{fig:context_mine}
\end{figure}

    \mypara{2. Contextual Relevance from Different Aspects} We observe that
    the context related to any given configuration line falls into one of
    three types, based on its position and relation within the
    configuration file as shown in the example in
    Figure~\ref{fig:context_mine}, with increasing levels of complexity to
    mine. They each providing unique insights that contribute to accurate
    misconfiguration detection: 
    
    \mysubpara{Neighboring Configurations} are closely related to the line under examination, typically within the same section or functional block of the network configuration. For instance, if analyzing a line that defines a VLAN assignment on a particular interface, neighboring configurations would include other lines that configure the same interface or VLAN. These configurations provide insights into how related parameters are set up in proximity, revealing potential dependencies or conflicts. Additionally, by referencing related configurations that define or modify these neighboring parameters, one can understand how changes in one part of the network might impact adjacent configurations, helping to identify misconfigurations that could lead to issues like traffic misrouting or security vulnerabilities. Neighboring configurations are relatively easy to mine because they are identified based on adjacency or location in the configuration file, making them straightforward to extract.
        
    \mysubpara{Similar Configurations} involve the same type of parameter or function as the configuration line under review but are applied in different contexts within the network. For example, consider configurations that assign IP addresses to different interfaces across various routers in the network. Even though the interfaces and routers may differ, the principles governing IP address assignment remain the same. By comparing these similar configurations, one can detect inconsistencies or deviations from standard practices that might indicate a misconfiguration. This type of context is essential for ensuring that configuration practices are consistent across the network, reducing the risk of errors that could lead to network outages or performance degradation. Mining these configurations is more challenging because, unlike neighboring configurations, they are not located adjacently but must be identified based on functional similarities.

    \mysubpara{Referenceable Configurations}
    provide essential definitions or additional information related to the
    parameter value being examined. In the network domain, this context is
    important for understanding how specific values are applied across different parts of the network configuration. For example, if a parameter value specifies an import policy, referenceable configurations might include other lines where this policy is defined or where its behavior is modified. By examining these configurations, one can gain a deeper understanding of how the policy influences routing decisions or interacts with other network elements, ensuring that the configuration is applied correctly and consistently throughout the network.
    Referenceable configurations are the most difficult to mine because they often involve indirect references and dependencies that are not immediately apparent, requiring a thorough and sometimes recursive analysis to trace how a value or policy is used across various parts of the configuration file.

Given the categorization of relevant contexts and the hierarchical tree model, we can automate the process of mining the aforementioned contexts by translating them into specific paths within the tree. This involves the following steps:

\paragraph{Neighboring Configuration Mining:}
    Following the hierarchical tree structure of the configuration file, we can model the entire configuration as a tree \( T \) with nodes representing configuration elements and edges depicting the relationships between them. The root node \(V_1\) represents the entry point of the configuration file and the \(V_2\) nodes represent the broadest configuration categories.
    For a given configuration path \( P = \{ V_1 \rightarrow V_2(P) \rightarrow V_3(P) \rightarrow \dots \rightarrow V_k(P) = v_k(P) \} \) within the tree, where \( v_k(P) \) is the parameter value assigned to the parameter node \( V_k(P) \), the goal of neighboring configuration mining is to extract other paths in the tree that share a common sub-path with \( P \) up to a certain depth. 
    
    The common ancestry level, or shared node depth, can be adjusted to control how much context is mined. Let us define \( m \) as the number of shared nodes from the root, then neighboring configurations are defined as:
\begin{multline*}
N_m(P) = \{ P' \in T \mid P' = V_1 \rightarrow  V_2(P')\dots \rightarrow V_m(P') \rightarrow \\ V_{m+1}(P') \rightarrow \dots \rightarrow V_k(P')= v_k(P') \},
\end{multline*}

where \( P' \) shares the first \( m \) nodes with \( P \), but the nodes following \( V_m(P') \) (denoted as \( V_{m+1}(P'), V_{m+2}(P'), \dots \)) can differ from the remaining nodes in \( P \), and 
\(P'\) does not end at the same parameter node as \(P\), \ie, \( V_k(P') \neq V_k(P) \).

Adjusting \( m \) allows for control over the size and relevance of the neighboring context.
A common choice is to set \(V_m(P) = V_m(P') = V_{k-1}(P) \), which captures sufficient context while avoiding unnecessary complexity. For example. let the configuration path under review be represented as
\[
P = \{V_1 \rightarrow A \rightarrow B \rightarrow C \rightarrow D = v_D \} \mid V_{k-1}(P) = C
\]
The set of neighboring paths \( N(P) \) can be defined as:
\[
N(P) = \{ P' \in T \mid P' = V_1 \rightarrow A \rightarrow B \rightarrow C \rightarrow D' = v_{D'}, D' \neq D \}
\]
For example, consider a configuration line that defines the IP address for a particular interface:
\begin{multline*}
P = \{V_1 \rightarrow \text{interfaces} \rightarrow \text{ge-0/0/0} 
\rightarrow \text{unit 0} \rightarrow \text{family inet}\\
\rightarrow \text{address} = 192.168.1.1/24 \}
\end{multline*}
Here, \(V_{k-1}(P)\) would be \( \text{`family inet'} \). Neighboring paths in this case might include:
\begin{multline*}
N(P) = \{V_1 \rightarrow \text{interfaces} \rightarrow \text{ge-0/0/0}
\rightarrow \text{unit 0} \rightarrow
\text{family}\\ \text{inet} \rightarrow \text{mtu} = 1500 \}
\end{multline*}
This ensures the extracted neighboring paths are closely relevant, revealing how related parameters are configured in the same section without overwhelming the system with excessive data. Thus, in our evaluation of \sysname{}, we select \(V_{k-1}(P)\) as  \( V_m(P) \) for an effective balance between relevance and computational efficiency.

\paragraph{Similar Configuration Mining:} To identify similar configurations, we focus on paths that share the same root node (\(V_1\)), broadest category node (\(V_2\)), and parameter node (\(V_k\)), but differ in their intermediate nodes or parameter values. Let \( P = \{ V_1 \rightarrow V_2(P) \rightarrow V_3(P) \rightarrow \dots \rightarrow V_k(P) = v_k(P) \} \) represent the configuration line under review again.
The set of similar configuration paths \( S(P) \) is defined as:
\begin{multline*}
S(P) = \{ P' : P' = \{ V_1 \rightarrow V_2(P') \rightarrow \dots \rightarrow V_k(P') = v_k(P') \} \},
\end{multline*}
where \( P' \) shares \(V_2(P') = V_2(P)\) and \(V_k(P') = V_k(P)\), but may have different intermediate nodes (\( V_3(P'), V_4(P'), \dots \)) and a different parameter value \( v_k(P') \).

By ensuring \(V_2(P') = V_2(P)\), we compare configurations within the same category or context, even if the paths leading to \( V_k(P') \) and \( V_k(P) \) differ. Using only the root node \( V_1 \) would be too broad, as it encompasses the entire configuration file and does not provide meaningful differentiation. This approach allows us to understand how the same type of parameter is configured across different instances, which is crucial for ensuring consistency in network configurations.

For example, let the configuration path under review be:
\[
P = \{ V_1 \rightarrow \text{Interfaces} \rightarrow \text{Ethernet0} \rightarrow \text{IP} \rightarrow \text{MTU} = 1500 \}
\]
This represents the configuration for setting the Maximum Transmission Unit (MTU) to 1500 on interface Ethernet0.
A similar configuration \( P' \) could be:
\[
P' = \{ V_1 \rightarrow \text{Interfaces} \rightarrow \text{Ethernet1} \rightarrow \text{IP} \rightarrow \text{MTU} = 9000 \}
\]
In this case, \( P' \) shares the same broad configuration category `Interfaces' (\( V_2(P') = V_2(P) \)) and parameter node `MTU' (\( V_k(P') = V_k(P) \)), but differs in the intermediate node `Ethernet1' and the parameter value (9000 vs. 1500). By comparing similar configurations, the context reveals not only the MTU settings but also the intermediate configuration elements leading to the MTU setting across different interfaces within the network configuration.

\paragraph{Referenceable Configuration Mining:} In the hierarchical tree structure, referenceable configurations are identified by locating paths where the parameter value of the current line under review appears as an intermediate node in other paths. This context is crucial for understanding how specific parameter values or policies are further referenced, elaborated upon, or applied in other parts of the configuration. we can formalize this set of referenceable configuration paths as:
\[
R(P) = \{ P' : P' =  V_1 \rightarrow \dots \rightarrow v_k(P) \rightarrow \dots \rightarrow V_k(P') \} = v_k(P').
\]
In these paths, \( v_k(P) \) is no longer a parameter value, but an intermediate node, meaning it is being referenced or defined further down the configuration hierarchy.

For example, suppose the current configuration line is:
\(V_1 \rightarrow RouterA \rightarrow Policy \rightarrow ImportPolicy = PolicyX\), a referenceable path might be:
\(
V_1 \rightarrow  RouterA \rightarrow Policy \rightarrow PolicyX \rightarrow Filter = AllowAll
\),
indicating that \( PolicyX \) is further applied or modified by the filter configuration. Referenceable configuration mining helps to understand not only how a particular configuration value is defined, but also how it interacts with or influences other components of the network. This process is essential for detecting misconfigurations that arise from improper application or dependency handling across different sections of the configuration file.


One of the most important problems in extracting referenceable configurations
is the risk of incorporating irrelevant context, particularly because the
process does not distinguish between pre-defined parameter values inherent to the configuration language and custom values defined by network administrators. 
We now proceed to discuss this problem in more detail and present our solution.

\subsubsection{Challenge 2 - Referenceable Parameter Value Ambiguity}

In network configurations, parameter values have two types: \textit{pre-defined} values and \textit{user-defined} values. Pre-defined values are those that are built into the configuration language itself, such as boolean flags (True vs. False) or access control decisions (Allow vs. Deny). These values are understood by the configuration parser and typically do not require any additional definition or context within the configuration file. On the other hand, user-defined values are customizable by the network administrator and can vary widely depending on the specific requirements of the network. These include values such as IP addresses, timeout intervals, VLAN IDs, and other numeric or alphanumeric identifiers. However, there rarely exists documentation explicit specifying these types and requires a lot of manual examination, which is hard to scale.

When extracting context for referenceable configurations,
failing to differentiate between pre-defined and user-defined values can introduce irrelevant or misleading information into the extracted context.
For instance, a pre-defined value like True is universally recognized by the configuration parser and could be used in multiple contexts, such as enabling a feature (FeatureX → Enabled = True) or setting a protocol flag (OSPF → PassiveInterface = True). Mining based on this pre-defined value could lead to irrelevant results, pulling in unrelated configurations that share the same boolean logic, thereby contaminating the context pool.
Conversely, user-defined values like IP addresses, a prefix limit (\eg, "maximum-prefixes 500" in a BGP configuration) or a timeout interval (\eg, 30 seconds) are specific to the network and typically set by network operators. These values may appear in routing tables, NAT rules, or timeout policies, with their significance depending on how they’re defined and applied. In this case, mining should focus on retrieving all related configurations that define or use these values.

Distinguishing between pre-defined and user-defined values is crucial to avoid extracting irrelevant context, which can lead to several issues:

\begin{enumerate}
    \item \textit{Context Contamination}: 
    Irrelevant context introduced during the context mining process might erroneously group together configurations that pertain to entirely different aspects of network operation.
    This contamination of the context pool dilutes the relevance of the mined information, reducing the precision of the LLM and increasing the likelihood of false positives or missed errors. 
    \item \textit{Reducing Computational Costs}: LLMs can be computationally expensive, especially when processing large-scale network configurations with many interdependent components. Distinguishing between pre and user-defined values can help optimizing the context extraction process, ensuring that only relevant and necessary contexts are passed to the model. This reduces the computational overhead associated with processing extraneous information.

\end{enumerate}

\mypara{Existence and Majority-Voting Based Differentiation}
To accurately differentiate between pre-defined and user-defined parameter values within configuration files, we propose a solution that combines existence checks within the configuration tree and a majority-voting mechanism.

\mysubpara{1. Existence-Based Differentiation:}
User-defined values often carry contextual information and are typically associated with further definitions or explanations elsewhere in the configuration file. For example, if a parameter value represents a specific, customized import policy, the configuration should contain other lines that elaborate on this policy's behavior. In the hierarchical tree model, if a parameter value is user-defined, it is likely to appear as an intermediate node in other configuration paths, indicating that it is referenced or elaborated upon elsewhere. Conversely, if no such paths exist, the value is likely pre-defined and requires no additional context. Consider the parameter RoutingPolicy with a value of ImportPolicy1 as an example. If ImportPolicy1 appears as an intermediate node in other configuration lines, such as those defining specific route maps or filters, it is likely user-defined. On the other hand, a parameter value like `PassiveInterface = True' might not have any additional references, indicating that True is a pre-defined value.

Formal Definitions:
Let \( \text{Val}_{\text{user}} \) be the set of user-defined values, and \( \text{Val}_{\text{pre}} \) be the set of pre-defined values.
A parameter value \( v_k \) is considered \textit{user-defined} if it appears as an intermediate node in at least one other configuration path \( P' \in T \). This can be expressed as:
\begin{multline*}
v_k \in \text{Val}_{\text{user}} \iff \exists P' = \{ V_1 \rightarrow \dots \rightarrow v_k \dots \rightarrow V_k(P') \\ = v_k(P') \} \in T
\end{multline*}

A parameter value \( v_k \) is considered \textit{pre-defined} if it does not appear as an intermediate node in any other configuration path \( P' \in T \). This can be formalized as:
\begin{multline*}
v_k \in \text{Val}_{\text{pre}} \iff \nexists P' = \{ V_1 \rightarrow \dots \rightarrow v_k \dots \rightarrow V_k(P') \\ = v_k(P') \} \in T
\end{multline*}


\mysubpara{2. Majority-Voting for Consistency:} A shortcoming with existence-based differentiation is that the same parameter value can be user-defined in one context and pre-defined in another. For instance, the value 1000 used in a timeout setting might be pre-defined and require no further explanation. However, the same value 1000 used as a policy name or group identifier could be user-defined and require contextual elaboration. This distinction is crucial, as identical values can have different meanings depending on the associated configuration parameter. To address this ambiguity, we avoid assigning a universal type to a parameter value across all configurations. Instead, we determine whether a value is pre-defined or user-defined based on the specific combination of the configuration parameter and its associated value. For each configuration parameter, we analyze all associated values. If the majority of these values are user-defined, we classify the entire parameter (\(V_k\)) as well as all of its possible values (\(v_k\)) as user-defined. Conversely, if the majority are pre-defined, the parameter is classified accordingly. This approach leverages the principle that configuration parameters should exhibit uniformity in the type of values they accept, maintaining consistency across the configuration.

This further transforms our previous definition of \( \text{Val}_{\text{user}} \)  and \( \text{Val}_{\text{pre}} \)  into:
\[
(V_k, v_k) = 
\begin{cases} 
\text{User-Defined}, \text{if } \sum_{i=1}^{n} \mathbb{1}(v_i \in \text{Val}_{\text{user}}) > \frac{n}{2}, \\
\text{Pre-Defined}, \text{if } \sum_{i=1}^{n} \mathbb{1}(v_i \in \text{Val}_{\text{pre}}) \geq \frac{n}{2}.
\end{cases}
\]
where \( n \) is the number of unique values associated with the configuration parameter \( V_k \), and \( \mathbb{1} \) is the indicator function that checks whether a value is user-defined or pre-defined.

Example: For the parameter Timeout, where values like 1000, 2000, and 3000 are used, if the majority lack associated definitions, they are treated as pre-defined. Conversely, for a parameter like ImportPolicy, where values such as PolicyA, PolicyB, and 1000 (as a policy name) are used, if most have contextual definitions, all values under that parameter are treated as user-defined.

By combining existence-based checks with a majority-voting mechanism, we can effectively differentiate between pre-defined and user-defined parameter values in network configurations. This method ensures that the context extracted for LLM analysis is both relevant and accurate, thereby enhancing performance and reducing computational costs. Moreover, this approach maintains consistency within the configuration, ensuring that similar parameters are uniformly treated across different contexts.

\subsection{Iterative Prompting}\label{prompting_method}
As we transition from context mining to utilizing the extracted information in LLM prompts, the next challenge is how to feed this information into the model without overwhelming it. While extracting accurate and relevant context is essential, the model's ability to process that information effectively is just as crucial.


\subsubsection{Challenge 3 - Context Overload in Prompting}
\label{challenge_3}
A significant limitation of many LLM-based approaches is context overload, where the model is fed too much information, causing it to lose focus on the most pertinent details.
Even with the precise context extraction mechanisms we have, the extracted context can sometimes be extensive,
leading to several issues:

\begin{enumerate}
    \item \textit{Dilution of Relevance}: The model might lose track of the most critical elements of the prompt, resulting in responses that are less accurate or less relevant. For instance, in a complex router configuration, key misconfiguration details could get buried in a flood of less relevant context.
    \item \textit{Loss of Coherence and Accuracy}:
    The model might produce disjointed responses or fail to address core
    misconfigurations when processing too much information at once. This
    can lead to fragmented reasoning or errors, especially with intricate
    router settings. Additionally, token limitations may cause the model
    to truncate important sections or struggle to process excessive context, 
    further impacting performance.
\end{enumerate}

For example, in detecting routing policy misconfigurations, a single misconfigured line may have dependencies across several sections of the file. Presenting all related context at once can cause the model to fail at prioritizing critical information, leading to missed/incorrect detection. STOA methods, such as prompt chaining, which guides the model through chained instructions, doesn't solve this issue, as it doesn't dynamically adjust or prioritize the context based on model needs.

\mypara{Solution}
To address these issues, \sysname{} implements an iterative, sequential prompting framework that allows the model to request and process relevant context in stages. Instead of overwhelming the model with all available context at once, \sysname{} enables the model to actively request additional information as needed. This approach enhances the model’s ability to maintain focus, coherence, and overall detection accuracy by allowing it to prioritize and process context more effectively, ensuring that only the most pertinent information is presented at each step.

\begin{figure}[tb]
    \centering
    \fbox{\includegraphics[width=0.95\columnwidth]{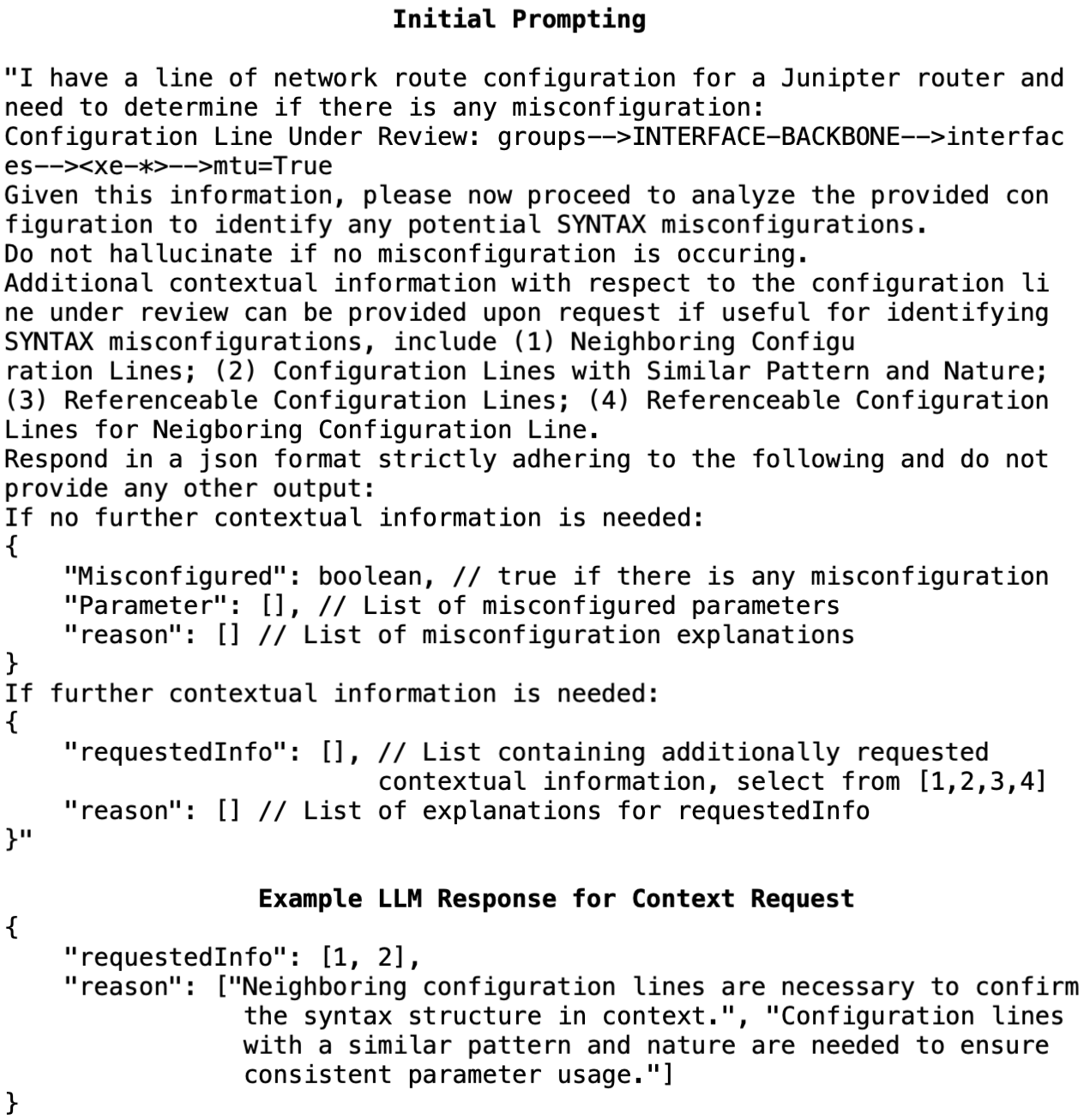}}
    \caption{Example: initial prompting and LLM context request response for detecting syntax misconfiguration.}
    \label{fig:initial_prompt}
\end{figure}

\mysubpara{1. Initial Prompting:} The process begins by feeding the LLM the specific configuration line under review, along with targeted instructions about the type of misconfiguration we are checking for (\eg, syntax errors, dependency conflicts, or just general misconfiguration).
Figure~\ref{fig:initial_prompt} presents an example where the initial prompt
includes MTU configuration line with an instruction to identify any
syntax misconfigurations related to that configuration.
This reduces the initial information load and allows the model to focus on the core issue.

\mysubpara{2. Contextual Options:} Next, \sysname{} offers the model the option to request additional context based on its understanding of the configuration line and the misconfiguration detection request. These options are based on our categorized context types, including: neighboring (\( N(P) \)), similar (\(S(P) \)) , and referenceable contexts (\( R(P) \)), as well as referenceable contexts on neighboring configurations (\( N(R(P)) \)). By allowing the model to choose which context to receive, \sysname{} ensures that the LLM is only given information that is most relevant to the detection task at hand, reducing the risk of context overload. Figure~\ref{fig:initial_prompt} illustrates an example request response
where the LLM requests neighboring context and referenceable contexts on neighboring configurations after the initial prompt to aid analyzing the MTU setting.
    
\begin{figure}[tb]
    \centering
    \fbox{\includegraphics[width=0.95\columnwidth]{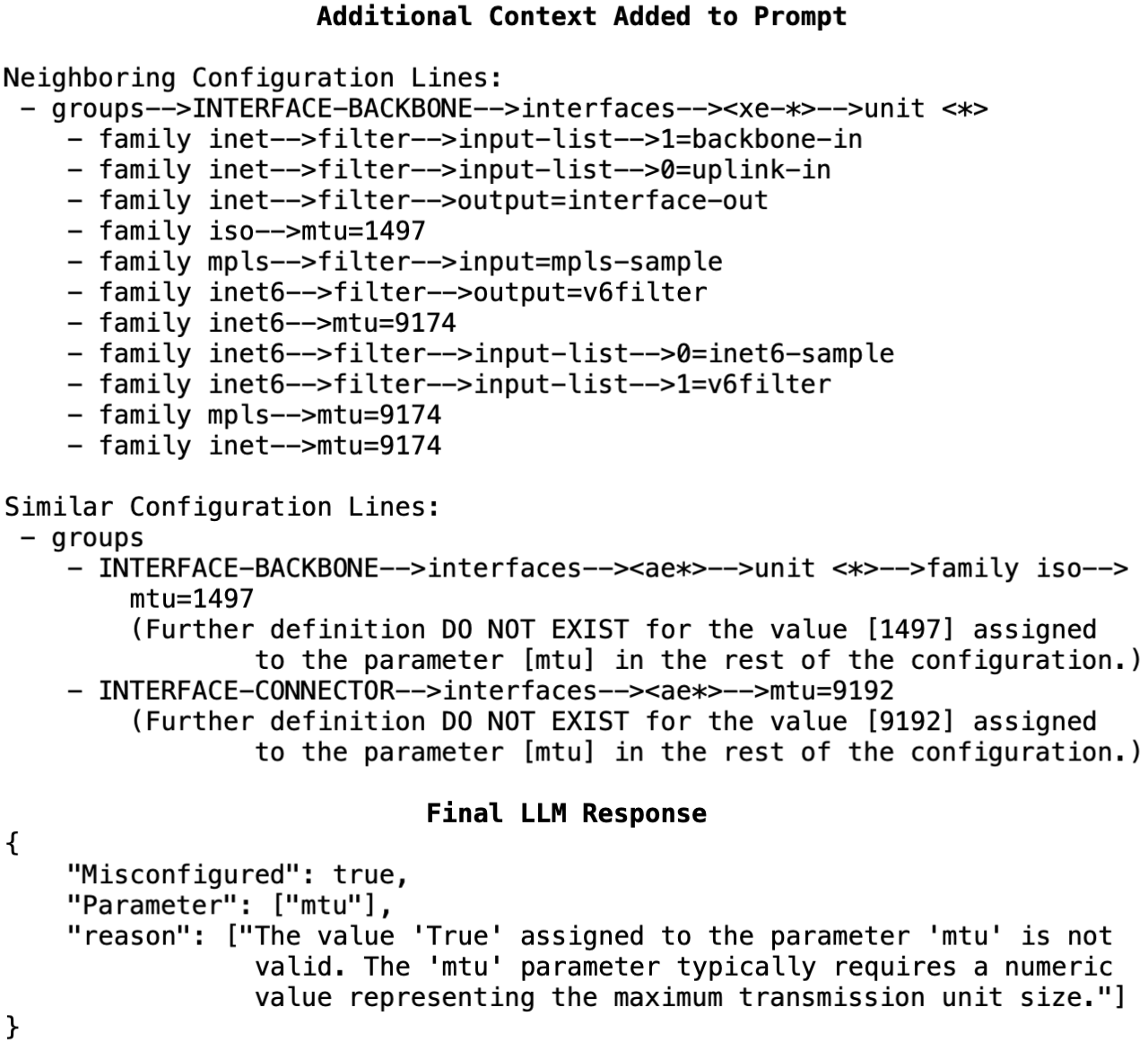}}
    \caption{Example: Adding requested context and retrieving misconfiguration detection result.}
    \label{fig:feedback_and_response}
\end{figure}

\mysubpara{3. Iterative Refinement:} After the model processes the initial prompt, it can request additional information if needed before arriving at the final decision. \sysname{} engages in a feedback loop where the model’s output informs the next set of prompts. For instance, if the model identifies a possible syntax issue but requires more context from a related policy definition, \ie, similar context, \sysname{} can deliver that specific context in the next prompt. This iterative process continues until the model has sufficient information to make a well-informed decision. Figure~\ref{fig:feedback_and_response} provides an example of what a final misconfiguration decision looks like regarding a misconfiguared MTU value.

This sequential, interactive approach mirrors how a network administrator might manually review a configuration file—examining the most relevant sections first, then digging deeper into referenceable or adjacent configurations as needed.


\subsubsection{Why Iterative, Requested Based Prompting Works}
By breaking down the context into smaller, more digestible portions, \sysname{} addresses the fundamental challenges of context overload. This method significantly enhances the model’s ability to:
\begin{enumerate}
    \item Stay on target: With less irrelevant information, the model can maintain focus on the specific issue under review.
    \item Preserve coherence: Incrementally adding information allows the model to build a more coherent understanding of the configuration~\cite{li2023prompt,subramonyam2024bridging}, improving detection of complex issues such as dependency conflicts or misapplied policies.
    \item Optimize performance: Avoiding context overload reduces the processing burden on the model, leading to faster and more reliable outputs.
\end{enumerate}

For example, in detecting misconfigurations in a multi-layer firewall policy, \sysname{} might first present the core rule in question, then iteratively offer neighboring rules that affect the same traffic flow, followed by referenceable configuration sections that define broader network policies. This structured process ensures that the LLM has all the information it needs, without drowning in unnecessary details, enabling it to make accurate decisions. Unlike conventional prompt-chaining, which provides incremental instructions based on a fixed context, \sysname{} dynamically adds context as requested by the LLM, focusing on the evolving needs of the analysis.


\section{Evaluation}
\label{sec:eval}
To evaluate the effectiveness of \sysname{}, we present two test scenarios as case studies and compare \sysname{} with state-of-the-art solutions. These include:

\begin{enumerate}
    \item \textit{Synthetic Misconfiguration Verification}: Synthetic misconfigurations are introduced into real network configuration snapshots. We run \sysname{} on specific configurations before and after the misconfigurations are introduced to evaluate the model’s ability to accurately infer and detect misconfigurations.
    \item \textit{Comprehensive Real Configuration Snapshot Verification}: \sysname{} is run on configuration snapshots from a medium-scale campus network to detect overlooked misconfigurations in a real-world setting. This covers both \textit{targeted} misconfiguration detection, where we focused on known issues like VLAN assignment errors, and \textit{non-targeted} detection, which involved scanning for general misconfigurations without prior knowledge of the issues.
\end{enumerate}

We now introduce the setup for our evaluation and present the details of the case studies in the subsequent sections.

\subsection{Evaluation Setup}

In this section, we describe our evaluation setup.

\subsubsection{Context Extraction Hardware Setup}

\textit{System Information}: GNU/Linux Red Hat Enterprise Linux release 8.8 (Ootpa); 
\textit{CPU Specifications}: Architecture: x86\_64; CPU(s): 32 (2 threads/core, 16 cores/socket, 1 socket); Model: AMD EPYC 7302P 16-Core Processor; Max MHz: 3000.000; L3 cache: 16384K; NUMA node0 CPU(s): 0-31 

We opt not to use the GPU, as the computational cost of the context extraction and processing tasks was manageable on the CPU --- the tasks primarily benefit from parallel CPU threads, and GPU acceleration would not provide significant additional speedup.

\subsubsection{LLM Model Used: GPT-4o} For running the LLM-based detection, we use GPT-4o~\cite{openai_gpt4o}, OpenAI's most recent transformer-based model designed for complex multi-step tasks.

\mypara{Specifications} Model: GPT-4o-2024-05-13; Context window: 128,000 tokens; Max output tokens: 4,096 tokens; Training data Up to October 2023.

\subsubsection{Prompting and Query Setup}
We use OpenAI’s Chat Completions API to interact with GPT-4o. The system was provided with a structured conversation history to maintain context across multiple queries. For each query, the model was tasked with analyzing the configuration lines to detect potential misconfigurations, with instructions tailored to focus on specific misconfiguration types (\eg, syntax issues, policy conflicts) or general misconfigurations. Figures~\ref{fig:initial_prompt} and ~\ref{fig:feedback_and_response}
show example prompting, queries, and responses.

\subsection{Case Study 1: Synthetic Misconfiguration Verification}
We first evaluate \sysname{} on synthetically introduced misconfigurations, which allows us to systematically test various misconfiguration types and assess detection accuracy in a controlled environment.
We broadly categorize router misconfigurations into three types:
\begin{itemize}
    \item \textit{Syntax errors} occur when the configuration does not adhere to the expected format or structure---e.g., a missing bracket or misused keyword in a BGP routing policy.
    \item \textit{Range violations} involve parameter values that fall outside the acceptable range---e.g., an MTU value that exceeds the maximum allowed for a specific interface type.
    \item \textit{Dependency/conflict (D/C) issues} arise when different configuration lines are incompatible or contradict each other---e.g., a firewall rule might block traffic that another policy explicitly permits.
\end{itemize}

\begin{table}[tb]
\centering
\resizebox{\columnwidth}{!}{
\begin{tabular}{|l|l|l|l|}
\hline
\textbf{Type} & \textbf{Description} & \textbf{Misconfig} & \textbf{Requested Context}\\ \hline

\multirow{4}{*}{\textbf{Syntax}} 
& Missing brace &...interfaces: \{\textcolor{red}{"<ge-*"}: & None\\ \cline{2-4}
& Invalid keyword & ...mtu: \textcolor{red}{"True"}   & \( N(P) \), \( R(P) \)\\ \cline{2-4}
& Incorrect hierarchy & ...host-name: \textcolor{red}{"\{rtsw.alba-re1\}"}  & \( N(P) \)\\ \cline{2-4}
& Invalid IP address & ...neighbor: \textcolor{red}{192.168.253.1.1}" & \( N(P) \), \( R(P) \)\\ \hline

\multirow{4}{*}{\textbf{Range}}  
& Invalid MTU & ...mtu: \textcolor{red}{"10000"}  & \( N(P) \), \( R(P) \)\\ \cline{2-4}
& Invalid VLAN ID & ...vlan-id: \textcolor{red}{"5000"}  & \( N(P) \), \( R(P) \), \( N(R(P)) \) \\ \cline{2-4}
& Invalid AS & ...autonomous-system: \textcolor{red}{70000} & \( N(P) \), \( R(P) \) \\ \cline{2-4}
& Invalid prefix limit & ...maximum: \textcolor{red}{"200000"} & \( N(P) \), \( R(P) \)  \\ \hline

\multirow{8}{*}{\textbf{D/C}}  
& Non-existent group & ... apply-groups: \textcolor{red}{"L2-LSP-ATTRIBUTES”} & \( N(P) \), \( R(P) \), \( N(R(P)) \) \\ \cline{2-4}
& Policy Conflict & ...community \textcolor{red}{add I2CLOUD-EXTENDED-TARGET}  & \( N(P) \), \( R(P) \), \( N(R(P)) \) \\ \cline{2-4}
& Non-existent filter & ...input-list: \textcolor{red}{"uplink"} & \( N(P) \), \( R(P) \)  \\ \cline{2-4}
& Non-existent policy & ...export: \textcolor{red}{"OESS-400-300-LOOP} & \( R(P) \)  \\ \cline{2-4}
& Incorrect Filter Usage & ...family mpls: filter:input-list:\textcolor{red}{"v6filter"} & \( R(P), S(P) \)  \\ \cline{2-4}
& Disabled Sampling & ...family inet6: \textcolor{red}{SAMPLING MISSING/DISABLED}& \( R(P), N(P) \)  \\ \cline{2-4}
& VRF Target Conflict & ...vrf-target target:11537:\textcolor{red}{"313001"}& \( S(P) \)  \\ \cline{2-4}
& Abnormal Small MTU & ...family iso: mtu: \textcolor{red}{"1497"}& \( R(P), S(P) \)  \\ \hline
\end{tabular}}
\caption{Synthetic Misconfigurations Introduced and LLM Requested Context for \sysname{}.}
\label{tab:syn_configs}
\end{table}

To evaluate \sysname{}, we obtain snapshot configurations from Internet2 (Juniper devices) and sampled 16 correct configuration lines. These were then modified to introduce synthetic errors across the three misconfiguration types. For syntax and range errors, we created four distinct misconfigurations each, totaling 8 errors (see Table~\ref{tab:syn_configs}). For D/C misconfigurations, we introduced 8 distinct errors, as vanilla LLMs excel at detecting syntax and range issues, while \sysname{}’s context mining is particularly effective at uncovering nuanced D/C issues. The additional D/C errors allow us to thoroughly assess this capability.
For each misconfiguration, we run \sysname{} by following the two components: (1) treating the misconfigured line as the line under review and extracting all relevant context, and (2) conducting the iterative, sequential prompting process against the model, obtaining the final misconfiguration detection decision.

When forming the prompt, we explicitly instruct the model to look for each type of misconfiguration—syntax, range, or dependency/conflict—individually. This is because the model may request different types of context depending on the specific misconfiguration type it is trying to detect. We report only the results corresponding to the actual misconfiguration type introduced. Importantly, \sysname{} never yielded false positives when the model was instructed to find a misconfiguration type different from the actual type. Additionally, we find that asking the model to search for `GENERAL' misconfigurations also led to successful detection, though more context was often requested by the model.

We compare \sysname{} to three representative tools: the model checker \textit{Batfish}~\cite{fogel2015general},  the consistency checker \textit{Diffy}~\cite{kakarla2024diffy}, and the partition-based GPT Q\&A model \textit{Ciri}~\cite{lian2023configuration} using GPT-4o for fair comparison. 

\begin{table}[thb]
\centering
\resizebox{\columnwidth}{!}{
\begin{tabular}{|l|l|l|l|l|l|}
\hline
\multirow{2}{*}{\textbf{Type}} & \multirow{2}{*}{\textbf{Cases}} & \multicolumn{4}{c|}{\textbf{Number of Corrected Detected Cases}} \\ \cline{3-6} 
&                                & \textbf{\sysname{}} & \textbf{Ciri} & \textbf{Batfish} & \textbf{Diffy} \\ \hline
Syntax Error & 4 & 4/4 (100\%) & 2/4 (50\%) & 3/4 (75\%) & 0/4 (0\%) \\ \hline
Range Error & 4 & 4/4 (100\%) & 2/4 (50\%) & 0/4 (0\%) & 0/4 (0\%) \\ \hline
D/C Error& 8 & 8/8 (100\%) & 1/8 (12.5\%) & 2/8 (25\%) & 0/8 (0\%) \\ \hline
\multirow{2}{*}{\makecell[{{l}}]{Original Configs\\ (No Error)}}& \multirow{2}{*}{\makecell[{{l}}]{16}} & \multirow{2}{*}{\makecell[{{l}}]{16/16 (100\%)}} & \multirow{2}{*}{\makecell[{{l}}]{16/16 (100\%)}} & \multirow{2}{*}{\makecell[{{l}}]{16/16 (100\%)}} & \multirow{2}{*}{\makecell[{{l}}]{16/16 (100\%)}} \\ 
&  &  &  &  &  \\ \hline
\textbf{Total} & \textbf{32} & \textbf{32/32 (100\%)} & \textbf{21/32 (65.6\%)} & \textbf{21/32 (65.6\%)} & \textbf{16/32 (50\%)} \\ \hline
\end{tabular}}
\caption{Synthetic Misconfiguration Detection Results.}
\label{tab:syn_result}
\end{table}

The results in Table~\ref{tab:syn_result} demonstrate that \sysname{} outperforms the other tools in detecting all three types of misconfigurations. It achieves a perfect 16/16 detection rate across the board for misconfigurations. This indicates that \sysname{}'s comprehensive context mining and iterative prompting process effectively identify errors in network configurations, even in more complex cases like D/C issues. In contrast, the other tools showed limitations, particularly when it came to detecting range violations and D/C issues.


\paragraph{Comparison Against Partition-based LLMs} The partition-based GPT model Ciri performs reasonably well in detecting syntax and range violations, with a detection rate of 50\% for both categories.
This success is due to the LLM model's training on vast, diverse text, including network-related documentation and configuration files, enabling it to recognize common syntax structures and numerical limits typically found in configurations. Transformer-based LLMs, with their self-attention mechanism, are particularly effective at identifying these common, localized issues.
However, it lags significantly behind \sysname{} in detecting D/C issues, with a 12.5\% detection rate in this category. The primary reason for this underperformance is the partitioning approach. This approach analyzes configuration sections in isolation, failing to capture the complex interdependencies between configuration lines, which is crucial for detecting D/C issues. The only D/C misconfiguration that Ciri successfully detects is the `VRF Target Conflict,' where the conflicting `route-distinguisher' value configuration happens to be in the same prompt partition, providing just enough context for accurate detection.

As shown in Table~\ref{tab:syn_configs} and~\ref{tab:syn_responses}, D/C issues, such as policy conflicts, require the model to understand and analyze interactions between different configuration elements.
\sysname{} addresses this by guiding the LLM to request the relevant contextual information it needs for accurate reasoning and detection. The example \sysname{}-enable LLM responses in Table~\ref{tab:syn_responses} clearly show that these contexts are being leveraged to derive the correct detection results. In contrast, partition-based GPT struggles to detect such misconfigurations effectively, as it analyzes configurations in isolation and can often miss these complex interdependencies.

\begin{table}[ht]
\centering
\resizebox{\columnwidth}{!}{
\begin{tabular}{|l|l|l|p{8cm}|} 
\hline
\textbf{Type} & \textbf{Misconfig} & \textbf{errParameter} & \textbf{Reason}\\ \hline

\multirow{3}{*}{\textbf{Syntax}} & \multirow{3}{*}{Missing brace} & \multirow{3}{*}{["<ge-*>"]} & "MTU value for 'family inet' is set to 9174 which is less than the interface MTU value of 9192. The MTU for 'family inet' should not be less than the interface MTU."\\\hline

\multirow{4}{*}{\textbf{Range}} & \multirow{4}{*}{Invalid VLAN ID} & \multirow{3}{*}{["maximum"]} & "The 'maximum' value of 200000 in the configuration line under review significantly exceeds the 'maximum' values of 1000 found in similar configuration lines, indicating a potential range misconfiguration."\\\hline

\multirow{20}{*}{\textbf{D/C}} & \multirow{5}{*}{Policy Conflict} & \multirow{5}{*}{["remove-comms-rt"]} & "There is a conflict in the configuration under term 'remove-comms-rt' where the action 'community add I2CLOUD-EXTENDED-TARGET' is followed by 'community delete I2CLOUD-EXTENDED-TARGET' which are contradictory actions."\\\cline{2-4}

 & \multirow{5}{*}{Disabled Sampling} & \multirow{5}{*}{["inet6"]} & "The family mpls is incorrectly using the IPv6 filter inet6-sample v6filter, which is intended for family inet6 in similar configuration lines. MPLS traffic should have its own specific filter, as the filtering requirements for MPLS are different from those for IPv6."\\\cline{2-4}
 & \multirow{5}{*}{VRF Target Conflict} & \multirow{5}{*}{["target"]} & "The route-distinguisher (RD) in neighoring configurations is set to 11537:381841, while the vrf-target is set to target:11537:313001. These values should typically align, or at least follow certain conventions for mapping VRFs to the correct target instances in MPLS/BGP L2VPN."\\\cline{2-4}
 & \multirow{5}{*}{Incorrect Filter Usage} & \multirow{5}{*}{["input-list"]} & "The family mpls is incorrectly using the IPv6 filter inet6-sample v6filter, which is intended for family inet6 in similar configuration lines. MPLS traffic should have its own specific filter, as the filtering requirements for MPLS are different from those for IPv6."\\\hline
\end{tabular}}
\caption{Example LLM Reasoning for \sysname{}.}
\label{tab:syn_responses}
\end{table}




\paragraph{Comparison Against Non-LLM-based Tools} Batfish performs well on
syntax errors (75\% detection rate) but struggles with range violations and
only partially detected D/C issues (2/8).  Batfish uses a rule-based model
checker, which excels in detecting predefined syntax errors but is less
effective in handling misconfigurations that involve contextual nuances, such
as range violations or dependency conflicts. Specifically, Batfish 
detects issues related to non-existent groups and policies (hard-coded into its
rule checks) but missed more subtle D/C misconfigurations, like policy
conflicts or filter dependencies.

Diffy performs poorly across all misconfigurations, with a 0/16 detection rate, due to its data-driven approach, which focuses on learning common usage patterns from configurations and flagging anomalies as potential bugs. While effective for detecting clear deviations from standard patterns in other configuration files,
complex issues like policy conflicts or range violations may not deviate from typical patterns but still be incorrect, which Diffy’s template-based anomaly detection cannot capture. Furthermore, Diffy's reliance on identifying deviations from learned patterns makes it less effective at detecting syntax errors and range violations. These types of misconfigurations often follow standard formatting and numerical values that may not stand out as anomalies, making them difficult for Diffy to detect without a granular understanding of configuration structure and constraints.

Finally, despite varying performance in detecting misconfigurations, all tools, including \sysname{}, Batfish, Diffy, and the partition-based GPT model Ciri, successfully marked all correctly configured lines as valid prior to the introduction of the synthetic errors, indicating strong capabilities in avoiding false positives for correctly functioning configurations. 

\subsection{Case Study 2: Comprehensive Real Configuration Snapshot Verification}
To evaluate \sysname{} on real-world misconfiguration detection, we obtain
Aruba router configuration snapshots from a medium-scale campus network. We
have two objectives: (1) to verify the scalability of \sysname{} when applied to larger network configurations, and (2) to investigate whether \sysname{} can detect potential misconfigurations that have not yet been identified by existing tools.

We perform two types of analysis, beginning with \textit{targeted misconfiguration detection:} A key aspect of this case study is the demonstration of \sysname{}'s flexibility in integrating additional context types under different scenarios. This is particularly useful when network operators have prior domain knowledge about specific misconfigurations they are trying to detect that fall outside of \sysname{}'s default context library.
To illustrate this flexibility, we consult the network operators who provided the configuration snapshots. They were specifically interested in detecting incorrect `VLAN assignments' --- a type of misconfiguration that often requires a network-wide view. Such misconfigurations are usually identifiable only when considering the configurations of other devices in the same network, as deviations from uniform configurations can indicate potential issues.
To address this, we introduce an additional, default context type, called `Intra-Router Consistency Context'.  This context type mines  the prevalence of the same parameter-value pair across other devices in the network, providing insights into whether a configuration is common or potentially erroneous. For this scenario, we modify the initial prompt to focus specifically on detecting `WRONG VLAN ASSIGNMENT'.

Example Intra-Router Consistency Context extracted:

\textit{`For the Configuration Line Under Review, the same configuration is found in 189 out of 191 other configuration files. (Significantly lower prevalence may indicate an uncommon or potentially erroneous configuration.)'}

The second analysis involves \textit{non-targeted misconfiguration detection}, where \sysname{} uses its default context library to aim the LLM for detecting general misconfigurations without prior knowledge of specific issues. In this mode, no specific misconfiguration type is provided to the model; instead, \sysname{} exhaustively analyzes the entire configuration file to identify `GENERAL' errors. 
This approach is particularly valuable for detecting overlooked misconfigurations in large, complex network environments where errors may not be immediately obvious or anticipated. By running \sysname{} in this mode, we assess its ability to detect subtle misconfigurations across the network without predefined expectations, ensuring it can operate effectively even when network operators are unsure of the specific issues they might face.

We perform exhaustive analysis using \sysname{} across 11 configuration files covering $\sim6\%$ of all devices on the campus network, applying the context mining framework to extract all relevant context for each configuration line. We then prompt the model to identify the corresponding misconfigurations, allowing it to dynamically request the necessary context during the iterative prompting phase. The detection results are presented in Table~\ref{tab:real_results}, highlighting \sysname{}'s ability to uncover new misconfigurations. For each of misconfigurations detected by \sysname{}, we verify the correctness of the inference decision with domain experts. 

\paragraph{Targeted Analysis} 
In the targeted analysis, \sysname{} demonstrates consistent performance
in identifying misconfigurations related to incorrect VLAN assignments. Out of the six misconfigurations detected, \sysname{} correctly flags two with a true positive rate (TPR) of 33.3\%. The identified misconfigurations involved wrong VLAN assignments that caused misrouting and access issues, a high-severity problem for network operations.

Despite the modest TPR, the false positives align with the domain experts' expectations. Network experts clarified the VLAN assignment deviations falsely flagged as misconfigurations are in fact intentional modifications tailored to the specific routers within the network. In this regard, we compare \sysname{}'s detection results (both true and false positives) with those from the experts' internal tool—a graph-based, non-LLM solution designed to check intra-router consistency— and the results are consistent.
Thus, while the false positives may appear significant, they represent deviations from standard configurations that would typically signal errors, but in these cases, are expected and intentional adjustments.

\begin{table}[tb]
\centering
\resizebox{\columnwidth}{!}{
\begin{tabular}{|l|l|l|l|l|}
\hline
\multirow{2}{*}{\makecell[{{l}}]{\textbf{Analysis} \\ \textbf{Type}}} & \multirow{2}{*}{\makecell{\textbf{Misconfigs}}} & \multirow{2}{*}{\makecell{\textbf{Count}}} &  \multirow{2}{*}{\makecell{\textbf{Severity (Reason)}}} &\multirow{2}{*}{\makecell[{{l}}]{\textbf{True Positive}\\ \textbf{Rate (TPR)}}}\\
&&&& \\\hline
\multirow{2}{*}{\makecell[{{l}}]{\textbf{Targeted}}} &\multirow{2}{*}{\makecell[{{l}}]{Wrong VLAN\\Assignment}} & \multirow{2}{*}{\makecell{6}} & 
\multirow{2}{*}{\makecell[{{l}}]{\textit{High}: Wrong VLANs permitted in configuration, causing\\access issues.}} & \multirow{2}{*}{\makecell{2/6 (33.3\%)}}\\ 
&&&& \\\hline
\multirow{8}{*}{\makecell[{{l}}]{Non-\\Targeted}} &\multirow{2}{*}{\makecell[{{l}}]{Invalid Subnet\\ Mask}} & \multirow{2}{*}{\makecell{2}} & 
\multirow{2}{*}{\makecell[{{l}}]{\textit{High}: Subnet mask usage, \eg, 255.255.253.0) \\violating binary boundary rules.}} 
& \multirow{2}{*}{\makecell{2/2 (100\%)}}\\ 
&&&& \\\cline{2-5}

 &\multirow{2}{*}{\makecell[{{l}}]{Inconsistent \\Policy Naming}} & \multirow{2}{*}{\makecell{14}} & 
\multirow{2}{*}{\makecell[{{l}}]{\textit{Low}: Misnamed `smnpread' for `snmp\\_communities',but with consistent usage.}} & \multirow{2}{*}{\makecell{14/14 (100\%)}}\\ 
&&&& \\\cline{2-5}
 &\multirow{2}{*}{\makecell[{{l}}]{Insecure SSH \\KEX Algorithm}} & \multirow{2}{*}{\makecell{1}} & 
\multirow{2}{*}{\makecell[{{l}}]{\textit{Low}: Insecure `diffie-hellman-group14-sha1' used\\(low risk as stronger algorithms are included).}} & \multirow{2}{*}{\makecell{1/1 (100\%)}}\\ 
&&&& \\\cline{2-5}
&\multirow{2}{*}{\makecell[{{l}}]{Ambiguous Rate\\Limit Values}} & \multirow{2}{*}{\makecell{2}} & 
\multirow{2}{*}{\makecell[{{l}}]{\textit{Ambiguous}: `Interval' and `burst' set to 0, leading\\ to nondeterministic behaviors.}} & \multirow{2}{*}{\makecell{2/2 (100\%)}}\\ 
&&&& \\\hline
\end{tabular}}
\caption{Comprehensive Real Configuration Snapshot Verification Results: \sysname{}-Detected Misconfigurations}
\label{tab:real_results}
\end{table}

\paragraph{Non-Targeted Analysis}
In the non-targeted analysis, \sysname{} was able to detect several types of misconfigurations across the provided configuration snapshots, as shown in Table~\ref{tab:real_results}.
The detected misconfigurations were categorized by severity based on their potential impact on network operations:
\begin{itemize}
    \item \textit{High Severity}: Misconfigurations in this category pose critical risks to the network and require immediate action --- \sysname{} identified invalid subnet mask (e.g., 255.255.253.0) usages that violate binary boundary rules, which could lead to routing issues and network instability.
    \item \textit{Low Severity}: These issues may not immediately disrupt network operations but should be addressed to ensure optimal functionality and avoid potential risks---
    \sysname{} identified many instances of misnamed configurations (\eg, `smnpread' instead of `snmpread' for `snmp\_communities') and inclusion of insecure SSH KEX algorithm which (`diffie-hellman-group14-sha1'), do not break the system but could lead to confusion and maintenance challenges.
    \item \textit{Ambiguous}: These misconfigurations are related to configurations that lead to undefined or nondeterministic behaviors --- \sysname{} found ambiguous rate limit values (`interval' and `burst' set to 0), which can cause unpredictable performance issues due to missing documentation on how these values should be handled, suggesting updates or further specifications in the documentation are needed.
\end{itemize}

Specifically, \sysname{} successfully identified 19 misconfigurations, including two cases of invalid subnet masks, 14 instances of inconsistent policy naming, one insecure SSH Key Exchange (KEX) algorithm, and two ambiguous rate limit values. Validated with domain experts, we find all misconfiguration cases are either valid or justifiable with a TPR of 100\% across all targeted misconfigurations.

Across synthetic and real-world configurations, \sysname{} consistently outperforms state-of-the-art tools like Batfish, Diffy, and partition-based GPT (Ciri), especially in detecting complex misconfigurations. \sysname{}'s dynamic context mining and iterative prompting allow it to capture interdependencies that other methods missed. Both in targeted and non-targeted scenarios, it demonstrates capability in identifying misconfigurations that traditional tools overlooked, making it a robust solution for network configuration validation.

\section{Discussion and Future Work}
\label{sec:future}

\sysname{} has proven effective in improving the detection of router misconfigurations, but there are several avenues for further exploration and refinement. One challenge is expanding the framework to better handle large-scale, distributed networks with interdependent routers. These environments often involve complex relationships between devices, and future work could focus on extending \sysname{} to capture and analyze multi-device contexts by default. Incorporating cross-router consistency checks and advanced network-wide policy validation could be valuable enhancements.

Another area for improvement is enhancing the LLM's ability to process real-time configuration changes. As networks evolve dynamically, incorporating real-time monitoring data into the context mining process would allow for continuous verification and faster misconfiguration detection. Integrating \sysname{} with network management tools that track operational metrics could provide valuable runtime data, further improving detection accuracy by correlating live data with configuration snapshots.

Future enhancements can also involve proposing and implementing automated fixes for detected misconfigurations with real-time updates. This would help network operators quickly address issues, streamlining troubleshooting and repairs.
\section{Conclusion}
\label{sec:conclusion}

In this paper, we introduced \sysname{}, a framework for router misconfiguration detection that leverages context-aware iterative prompting to improve LLM accuracy. By systematically extracting relevant context from network configurations
and following a guided, interactive prompting mechanism, \sysname{} addresses limitations in current model checkers, consistency checkers, and LLM-based models. Through case studies involving both synthetic and real-world misconfigurations, we demonstrated that \sysname{} outperforms existing methods across different misconfiguration types, ensuring more reliable and scalable network configuration verification.
\label{endOfBody}
\bibliographystyle{plain}
\bibliography{citations} 
\end{sloppypar}
\label{LastPage}

\pagebreak
\end{document}